# Polar phonon anomalies in single crystalline TbScO$_3$

Stanislav Kamba,[a,*] Veronica Goian,[a] Dmitry Nuzhnyy,[a] Viktor Bovtun,[a] Martin Kempa,[a] Jan Prokleška,[b] Margitta Bernhagen,[c] Reinhard Uecker,[c] and Darrell G. Schlom[d,e]

[a]*Institute of Physics, Academy of Sciences of the Czech Republic, Na Slovance 2, 182 21 Prague 8, Czech Republic*
[b]*Faculty of Mathematics and Physics, Department of Condensed Matter Physics, Charles University, Ke Karlovu 5, 121 16, Prague 2, Czech Republic*
[c]*Leibniz Institute for Crystal Growth, Max-Born-Strasse 2, 12489 Berlin, Germany*
[d]*Department of Materials Science and Engineering, Cornell University, Ithaca, New York, 14853-1501, USA*
[e]*Kavli Institute at Cornell for Nanoscale Science, Ithaca, New York, 14853-1501, USA*

*Corresponding author. Email: kamba@fzu.cz

Polarized infrared reflectivity spectra of a (110)-oriented TbScO$_3$ single crystal plate were measured down to 10 K. The number of observed polar phonons active along the crystallographic $c$ axis at low temperatures is much higher than predicted by factor-group analysis for the orthorhombic $Pbnm$- $D_{2h}^{16}$ space group. Moreover, the lowest frequency phonons active in $E \parallel c$ as well as in $E \parallel [1\bar{1}0]$ polarized spectra exhibit dramatic softening tending to a lattice instability at low temperatures. The dielectric permittivity at microwave frequencies does not show any ferroelectric-like anomaly, but the dielectric loss exhibits a maximum at 100 K. The origin of the discrepancy between the number of predicted and observed polar phonons as well as the tendency toward lattice instability are discussed. Magnetic measurements reveal an antiferromagnetic phase transition near 3 K.

I. Introduction

Rare-earth scandates ($RE$ScO$_3$, with $RE$ = rare earth) have been intensively studied in the last decade in both single crystalline[1,2,3,4,5,6] and thin film[7] form. They have rather large dielectric constants ($K$=20-30)[8] and are chemically stable in contact with Si,[9] which makes them potentially promising as high-$K$ dielectrics for field-effect transistor applications.[9,10]

Another interesting application is in the field of photonic crystals and metamaterials. For example, eutectic $Tb_3Sc_2Al_3O_{12}$-$TbScO_3$ crystals make self-organized dielectric microstructure of perovskite fibers embedded in a garnet phase matrix.[11] Both phases can be etched away, depending on the composition, leaving a pseudo-hexagonally packed dielectric array of pillars or an array of pseudo-hexagonally packed holes in the dielectric material. Both structures can be filled with a metal or another material and, hence, they have possible applications as metamaterials or photonic crystals.[11]

The third and probably now the most frequent application of single-crystalline rare-earth scandates, is their usage as substrates for the epitaxial growth of high-quality perovskite-type thin films.[12] Such high quality films allow strain engineering of ferroelectric and multiferroic properties by choosing from different rare-earth scandates with different lattice constants to grow strained films upon.[13] For example, $SrTiO_3$ is a quantum paraelectric in the bulk form, but 1% tensile strained $SrTiO_3$ thin films deposited on $DyScO_3$ exhibit ferroelectric behavior below 270 K.[14] In a similar way, simultaneous ferromagnetic and ferroelectric order can be induced in thin films by straining $EuTiO_3$, which in its unstrained state is antiferromagnetic and a quantum paraelectric. This was achieved on $DyScO_3$ substrates, which induce +1% tensile strain in the $EuTiO_3$ thin films.[15] Both compressive and tensile strain can remarkably enhance the ferroelectric phase transition temperature $T_C$ (e.g., in $BaTiO_3$)[16].

It is prudent to study strained ferroelectric thin films that are very thin, below the critical thickness for relaxation, to keep the films commensurate and the strain homogeneous. This is because the dislocations introduced during relaxation are detrimental to ferroelectricity.[17,18,19] As a consequence, the typical thickness of such films is less than 50 nm. Direct dielectric measurements are rather difficult due to the leakage current frequently present in such films, therefore spectroscopic methods are frequently used.

By using THz and infrared (IR) spectroscopies it is possible to reveal the ferroelectric soft modes, which drive displacive phase transitions. Both spectroscopies were used with success for the study of the dynamics of the phase transition as well as the tuning of phonons by using an external electric field in $SrTiO_3$ thin films deposited on $DyScO_3$ substrates.[20,21,22] IR spectroscopy enabled the determination of the ferroelectric phase transition in a strained $EuTiO_3$ deposited on a $DyScO_3$ substrate.[15] For these kinds of experiments it is necessary first to characterize the phonons in the bare substrate at relevant temperatures and then the substrate phonon parameters can be used for the evaluation of the thin film spectra.

Phonons in the rare-earth scandates have been investigated using first principles calculations,[8,23] but detailed experimental data are missing. DyScO$_3$ has been studied using IR spectroscopy, but Baldassare et al.[24] published only unpolarized IR reflectivity spectra and assigned the modes solely on the basis of the calculated phonon frequencies in Ref. 23. Polarized IR reflectivity spectra of DyScO$_3$ were published in the supplement of Ref. 15, but without complete phonon parameters. Detailed Raman scattering spectra of DyScO$_3$ and GdScO$_3$ were published by Chaix-Pluchery and Kreisel.[25]

TbScO$_3$ is isostructural with and has similar lattice parameters to DyScO$_3$. Unpolarized room-temperature far-IR reflectivity spectra of TbScO$_3$ were briefly published, but no assignment of the polar phonons has been done.[26] The aim of this paper is to study the polarized IR reflectivity spectra of TbScO$_3$ down to 10 K. Rather large phonon frequency shifts with temperature were discovered. Their origin is discussed. In addition, we have investigated the magnetic properties of TbScO$_3$ and observe an antiferromagnetic phase transition near 3 K.

**II. Experimental**

A single-crystalline plate of $(1\bar{1}0)$ TbScO$_3$ with size 10x10x1 mm was utilized in this study. The crystal growth is described in detail by Uecker et al.[4] The crystal orientation allowed us to perform polarized IR and THz experiments with the polarization of the electric vector of the beam in the $\boldsymbol{E} \parallel [1\bar{1}0]$ and $\boldsymbol{E} \parallel [001]$ directions. It should be stressed that we use the non-standard setting with space group *Pbnm* to describe the crystallography of TbScO$_3$. In this setting its axes lengths are a = 5.4543 Å, b = 5.7233 Å and c = 7.9147 Å The structure determination of Veličkov et al. makes use of the standard setting of space group *Pnma* for the structure of TbScO$_3$ in the settings with c<a<b.[27]

For the THz time-domain transmission experiments we used a Ti:sapphire femtosecond laser oscillator. Linearly polarized THz probing pulses were generated by an interdigitated photoconducting GaAs switch and detected using the electro-optic sampling with a 1 mm thick [110] ZnTe crystal. The complex dielectric spectra were taken in the range of 5-66 cm$^{-1}$ (150 GHz – 2 THz) at temperatures from 10 to 300 K.

Near-normal IR reflectivity spectra were obtained using a Fourier transform IR spectrometer Bruker IFS 113v. In both THz and IR measurements, an Optistat CF cryostat (Oxford Instruments) with polyethylene (IR) and Mylar (THz) windows was used for measurements between 10 and 300 K. The frequency range of the low-temperature IR measurements was limited by the transparency of the polyethylene windows (up to 650 cm$^{-1}$); the measurements above room temperature were performed up to 3000 cm$^{-1}$, which allowed us to determine not only phonon parameters, but also $\varepsilon_\infty$, i.e. the sum of electronic contributions to the permittivity.

The near-normal IR reflectivity can be expressed as

$$R(\omega) = \left| \frac{\sqrt{\varepsilon^*(\omega)} - 1}{\sqrt{\varepsilon^*(\omega)} + 1} \right|^2 , \qquad (1)$$

where the complex dielectric function $\varepsilon^*(\omega)$ can be expressed using a generalized oscillator model with factorized form of the complex permittivity:

$$\varepsilon^*(\omega) = \varepsilon'(\omega) - i\varepsilon''(\omega) = \varepsilon_\infty \prod_j \frac{\omega_{LOj}^2 - \omega^2 + i\omega\gamma_{LOj}}{\omega_{TOj}^2 - \omega^2 + i\omega\gamma_{TOj}} \qquad (2)$$

where $\omega_{TOj}$ and $\omega_{LOj}$ are the transverse and longitudinal frequencies of the j$^{th}$ polar phonon, respectively, $\gamma_{TOj}$ and $\gamma_{LOj}$ are their damping constants, and $\varepsilon_\infty$ denotes the high frequency permittivity resulting from electronic absorption processes. The static permittivity $\varepsilon(0)$ is given by the sum of phonon and electronic contributions

$$\varepsilon(0) = \sum_{j=1}^{n} \Delta\varepsilon_j + \varepsilon_\infty . \qquad (3)$$

Here the dielectric strength $\Delta\varepsilon_j$ of the j$^{th}$ polar phonon is defined as

$$\Delta\varepsilon_j = \frac{\varepsilon_\infty}{\omega_{TOj}^2} \frac{\prod_k (\omega_{LOk}^2 - \omega_{TOj}^2)}{\prod_{k \neq j} (\omega_{TOk}^2 - \omega_{TOj}^2)} . \qquad (4)$$

Equations (1) – (2) were used for simultaneous fits of the IR reflectivity and complex THz permittivity; Equations (3) and (4) were used to determine the static permittivity.

The dielectric properties of the TbScO$_3$ plate at microwave frequencies were characterized by a thin dielectric resonator method.[28] A TE$_{01\delta}$ resonance of the 0.486 mm thick TbScO$_3$ plate was excited in a cylindrical shielding cavity with low coupling. The temperature dependence of the resonance frequency and quality factor was measured using an Agilent E8364B network analyzer and a Janis closed-circle He cryostat. Then the temperature dependence of the in-plane averaged dielectric parameters were calculated.[28]

Magnetic susceptibility and magnetization measurements were performed using a Quantum Design PPMS9 equipped with a home-made induction coil that enables measurements of *ac* magnetic susceptibility, $\chi$, from 13 Hz to 10 kHz at temperatures between 1.8 and 300 K.

### III. Results and discussion

The result of magnetic susceptibility measurements is shown in Figure 1. Susceptibility follows the Curie-Weiss Law $\chi(T) = \frac{C}{T-\theta}$ (see the solid line in Figure 1) with an extrapolated temperature θ = −5.1 K. The observed temperature dependence of the susceptibility with a peak near 3.0 K is typical of antiferromagnetic (AFM) materials. The magnetic properties of TbScO$_3$ were never investigated before, but DyScO$_3$ exhibits similar behavior, because its Néel temperature is 3.1 K.[29] DyScO$_3$ shows a strong magnetic anisotropy with an easy axis along the [100] direction and a hard axis along the [001] direction.[29] Our measurements were performed in the $[1\bar{1}0]$ direction, i.e., perpendicular to the magnetically hard axis. In DyScO$_3$ rather strong frequency dependence of magnetic susceptibility was reported; the peak in $\chi$(T) splits and the higher-temperature peak shifts to higher temperature with increasing frequency. Such behavior is typical of a spin glass or spin ice.[29] In contrast to DyScO$_3$, the susceptibility in TbScO$_3$ exhibits no frequency dependence. We also performed field-cooled susceptibility measurements in 0.1 T (not shown) and found them to be the same as zero-field cooled measurements, i.e., our system does not exhibit any spin glass behavior. Additionally, we performed magnetization measurements M(H) at 1.8 K (not shown). The magnetization quickly increases with magnetic field H, starts to saturate above 1 T, and reaches 110 emu/g at 5 T.

Polarized IR reflectivity spectra plotted at various temperatures are shown in Figure 2. The spectra above 150 cm$^{-1}$ show no dramatic temperature dependence. Phonon damping classically decreases on cooling and therefore the intensity of reflection bands increases on cooling. Non-trivial and unusual temperature dependence is, however, seen at low frequencies in both of the polarized spectra. For example, in the **E** ∥ $[1\bar{1}0]$ spectra the lowest frequency phonon softens and its damping increases on cooling down to 100 K. This is manifested in the **E** ∥ $[1\bar{1}0]$ reflectivity spectra by the shift-down of the lowest frequency reflection band and by a decrease of its strength on cooling to 100 K. Nevertheless, on further cooling to 10 K the phonon damping decreases and therefore the strength of this mode again increases (see Figure 2(a)). This temperature behavior is also clearly seen in the complex dielectric spectra plotted in Figure 3(a) as well as in Figure 4(a), where the temperature dependence of the two lowest phonon frequencies is shown. It follows from the sum rule that the oscillator strength $f_j = \Delta\varepsilon_j \omega_{TOj}^2$ of uncoupled phonons is usually temperature independent. The TO1 phonon frequency softens on cooling, and therefore also the dielectric strength $\Delta\varepsilon_1$ of this mode increases on cooling (see Figure 4). $\Delta\varepsilon_1$ increases, however, faster on cooling than follows from the temperature independent oscillator strength. This is caused by a coupling with the TO2 mode, which partially transfers its strength $\Delta\varepsilon_2$ to $\Delta\varepsilon_1$.

More complicated phonon behavior with temperature is seen in the **E** ∥ [001] polarized spectra (see Figures 2b, 3b and 5). One can also see that the broad reflection band between 100 and 180 cm$^{-1}$ splits on cooling below 100 K. A detailed fit of reflectivity spectra in Figure 2 reveals that no new phonon activates in the spectra in this frequency range; all modes are active up to room temperature, but the strength and damping of some of them exhibit strong temperature dependences. This provides evidence of the mutual coupling among the modes.

Before we discuss the origin of the detected phonon anomalies with temperature, let us compare the number of observed phonons with the factor group analysis. TbScO$_3$ at room temperature has an orthorhombic *Pbnm* perovskite structure with 4 formula units per unit cell.[27] (note that again we use crystal axes settings where a<b<c,[8] while Ref. 27 uses settings with c<a<b, where the space group is *Pnma*). Factor group analysis of the zone center phonons gives

$$\Gamma = 7A_g(x^2) + 8A_u(-) + 7B_{1g}(xy) + 5B_{2g}(xz) + 5B_{3g}(yz) + 8B_{1u}(z) + 10B_{2u}(y) + 10B_{3u}(x) \quad (5)$$

Symbols in brackets mark the activity of phonons in the IR and Raman spectra. It means that 8A$_u$ symmetry modes are silent (non-active), 24 phonons should be Raman active and 25 should be IR active (1B$_{1u}$+1B$_{2u}$+1B$_{3u}$ are acoustic phonons). According to this analysis 7 B$_{1u}$ phonons should be IR active in the **E** ∥ [001] spectra, while 18 modes (9B$_{2u}$+9B$_{3u}$) can be active in the ***E*** ∥ $[1\bar{1}0]$ polarized spectra. Parameters of all of the phonons observed in the spectra at 10 K are summarized in Table I. In the ***E*** ∥ $[1\bar{1}0]$ polarized spectra we observed 17 of the 18 allowed modes. This is in very good agreement with the theory, that the missing mode can be overlapped by some other mode. However, in the **E** ∥ [001] spectra we see 15 polar modes instead of the 7$B_{1u}$ symmetry phonons expected. This is a rather large discrepancy. In Figure 2b, 7 distinct reflection bands are seen, but a careful fit of the **E** ∥ [001] spectra reveal an additional 8 modes are present in the low-frequency range below 180 cm$^{-1}$ (with the exception of the weak mode at 490 cm$^{-1}$). There are several possible explanations of this fact:

1) The additional modes are leakage from the $B_{2u}$ and $B_{3u}$ symmetry modes, which are seen in the **E** ∥ [001] spectra due to the non-perfect crystal orientation or non-ideal polarization of the IR beam. In this case the additional weak modes should have counter-parts in the ***E*** ∥ $[1\bar{1}0]$ spectra (and vice-versa). Unfortunately, the frequencies of the $B_{2u}$ and $B_{3u}$ symmetry modes cannot be exactly distinguished and their frequencies can be partially influenced by overlapping of both symmetry modes. Anyway, one can see that most of the modes have no counter-parts in the different symmetry spectra.

2) The additional modes are not phonons, but electron transitions. Such excitations are frequently seen in rare-earth compounds, but their frequencies are usually not temperature dependent. In our case the low-frequency modes exhibit rather large temperature dependences.

3) The sample has a different symmetry than the *Pbnm* space group reported up to now or the unit cell contains more than 4 formula units.[27] Phonon frequency changes with temperature indicate some tendency toward lattice instability in TbScO$_3$ (note the maximum of the damping of the TO1 phonon seen near 100 K in the ***E*** ∥ $[1\bar{1}0]$ spectra and continuous softening of the TO1 modes on cooling in both polarized spectra – see Figures 4a and 5), but no clear new modes are observed to activate in the spectra on cooling. Practically all of the modes are active at all temperatures below 300 K. These observations provide no evidence for any structural change on cooling. Most of the additional modes are seen in the low frequency region below 180 cm$^{-1}$. If we assume multiplication of the unit cell (in principle already near room temperature), Brillouin zone (BZ) folding occurs and some new phonons can activate in the IR spectra. High-frequency phonon dispersion branches usually exhibit small dispersion in the BZ, therefore the newly activated modes from the BZ boundary have frequencies very close to the phonons from the BZ center and therefore they are hardly distinguishable in the spectra. On the other hand, the low-frequency phonon branches exhibit usually larger dispersion and the phonons from the BZ edge can have completely different frequencies than the modes from the BZ center. Also the acoustic mode from the BZ edge can become IR active after the BZ folding; these could be the lowest-frequency modes observed in both polarized spectra. Up to now no low-temperature structural studies have been performed, so it is not known whether the TbScO$_3$ crystal undergoes any structural phase transition. Microwave dielectric permittivity exhibits no ferroelectric-like anomaly, it just continuously increases on cooling due to the phonon softening. On the other hand, the dielectric loss exhibits a maximum near 100 K (see Figure 4b), i.e., the temperature where the TO1 mode in the $\boldsymbol{E} \parallel [1\bar{1}0]$ spectra exhibits maximal damping. A maximum in dielectric loss can be a signature of some structural phase transition, but it can also originate from some defects in the lattice.

It is worth noting that in the $\boldsymbol{E} \parallel [001]$ polarization the DyScO$_3$ and GdScO$_3$ crystals exhibit similar IR spectra as TbScO$_3$, but their phonon frequencies are slightly shifted because of the different masses of *RE* cations.[30] The reflection band between 100 and 180 cm$^{-1}$ exhibits smaller splitting than in TbScO$_3$, but in all samples it can be fit with more than one oscillator. In all of the above mentioned crystals more than the 7B$_{1u}$ modes are seen. It should be also stressed that first-principles phonon calculations performed for all of the scandates never gave phonon frequencies below 95 cm$^{-1}$.[8,23] On the other hand, in *RE*ScO$_3$ (*RE* = Dy, Gd, Tb) crystals we always found a B$_{1u}$ phonon with a frequency around 40 cm$^{-1}$. This shows either that the calculations are not accurate or the crystal structure of *RE*ScO$_3$ is different (presumably with a larger unit cell).

## IV. Conclusion

An antiferromagnetic phase transition has been discovered to occur in single crystalline TbScO$_3$ near 3 K. IR reflectivity spectra studied in both in-plane polarizations reveal softening of the lowest-frequency phonons on cooling. Moreover, the TO1 phonon in spectra with $\boldsymbol{E} \parallel [1\bar{1}0]$ exhibits an unusual increase of damping on cooling with a maximum near 100 K. We have performed factor group analysis of the BZ center phonons and compared them with our IR spectra. 15 phonons rather than the 7B$_{1u}$ symmetry allowed modes have been observed in spectra with $\boldsymbol{E} \parallel [001]$. The activation of additional modes could be explained by leakage from other symmetry modes, but more probably by a lower crystal symmetry or a larger unit cell. Clarification of this issue requires additional structural studies at low temperature, which are in progress.

**Acknowledgements**

This work was supported by the Czech Science Foundation (Project No. P204/12/1163) and MŠMT (COST MP0904 project LD12026). We thank J. Drahokoupil for his X-ray diffraction measurements, Xianglin Ke for helpful discussions, and J. Petzelt for his careful reading of the manuscript.

Table I. Phonon parameters obtained from the fits of polarized reflectivity spectra taken at 10 K. $\varepsilon_\infty$=4.2.

| | $E \parallel [001]$ ; $B_{1u}$ symmetry modes | | | | | $E \parallel [1\bar{1}0]$ ; $B_{2u}+B_{3u}$ symmetry modes | | | | |
|---|---|---|---|---|---|---|---|---|---|---|
| No. | $\omega_{TO}$ (cm$^{-1}$) | $\gamma_{TO}$ (cm$^{-1}$) | $\omega_{LO}$ (cm$^{-1}$) | $\gamma_{LO}$ (cm$^{-1}$) | $\Delta\varepsilon$ | $\omega_{TO}$ (cm$^{-1}$) | $\gamma_{TO}$ (cm$^{-1}$) | $\omega_{LO}$ (cm$^{-1}$) | $\gamma_{LO}$ (cm$^{-1}$) | $\Delta\varepsilon$ |
| 1 | 41.6 | 8.6 | 43.4 | 6.8 | 3.2 | 60.9 | 30.8 | 61.5 | 32.9 | 0.5 |
| 2 | 80.3 | 37.5 | 85.5 | 38.3 | 5.9 | 81.0 | 8.2 | 86.4 | 6.6 | 3.0 |
| 3 | 88.8 | 19.2 | 89.5 | 17.95 | 0.3 | 90.0 | 6.9 | 91.0 | 7.1 | 0.2 |
| 4 | 104.7 | 9.9 | 112.3 | 14.6 | 5 | 115.2 | 8.2 | 115.7 | 7.3 | 0.2 |
| 5 | 122.5 | 10.8 | 122.8 | 35.6 | 0.25 | 128 | 19.6 | 128.2 | 19.5 | 0.05 |
| 6 | 132 | 12.1 | 142 | 14.4 | 5.9 | 192.5 | 8.5 | 219.0 | 8.8 | 5.5 |
| 7 | 143.9 | 10.3 | 144.8 | 18.1 | 0.3 | 225.3 | 10.6 | 226.3 | 9.1 | 0.04 |
| 8 | 147.8 | 41.2 | 180.7 | 5 | 1.9 | 258.5 | 9.4 | 263.3 | 9.7 | 0.6 |
| 9 | 198.7 | 2.4 | 200.4 | 0.2 | 12.6 | 296.5 | 1.9 | 304.0 | 2.8 | 1.3 |
| 10 | 204.5 | 43.5 | 209.0 | 48.9 | 0.2 | 317.8 | 5.8 | 329.6 | 8.9 | 1.1 |
| 11 | 304 | 1.2 | 328.9 | 4.0 | 2.9 | 346.2 | 2.9 | 348.9 | 2.7 | 0.3 |
| 12 | 353.8 | 2.3 | 362.8 | 4.3 | 0.7 | 367.6 | 7.2 | 370.4 | 8.0 | 0.6 |
| 13 | 392.5 | 3.81 | 476.8 | 2.9 | 2.7 | 382.2 | 4.3 | 422.9 | 4.0 | 2.5 |
| 14 | 489.9 | 4.7 | 491.3 | 6.9 | 0.04 | 444.3 | 3.3 | 492.7 | 4.9 | 1.2 |
| 15 | 503.1 | 4.4 | 656.5 | 14.9 | 0.5 | 544.5 | 4.3 | 562.8 | 4.9 | 0.2 |
| 16 | | | | | | 565.3 | 5.2 | 691.0 | 19.5 | 0.06 |

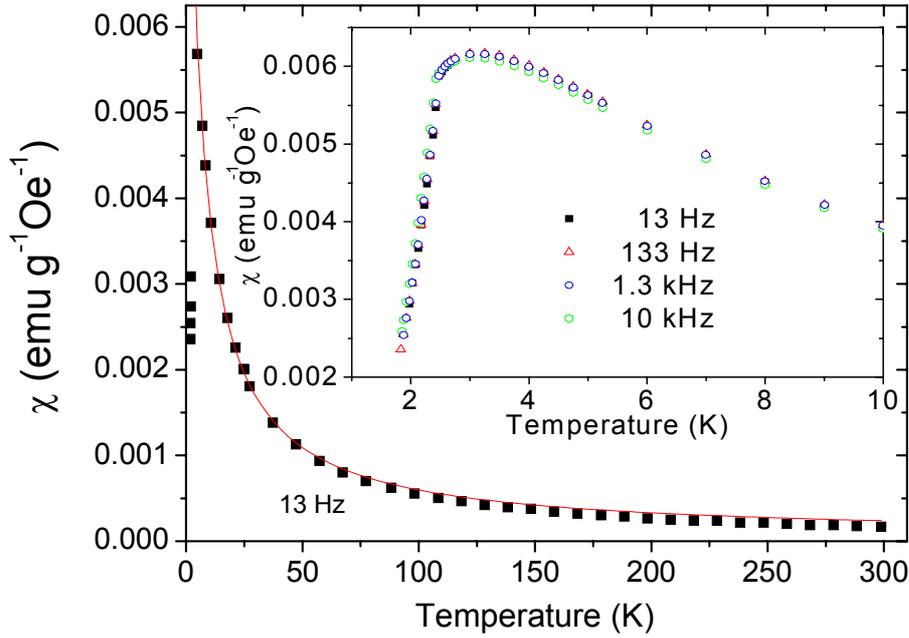

Figure 1. Temperature dependence of the *ac* magnetic susceptibility measured along the $[1\bar{1}0]$ direction on cooling, showing an AFM anomaly near 3 K. The line is the result of the Curie-Weiss fit. The inset shows the low temperature data taken at various frequencies on heating.

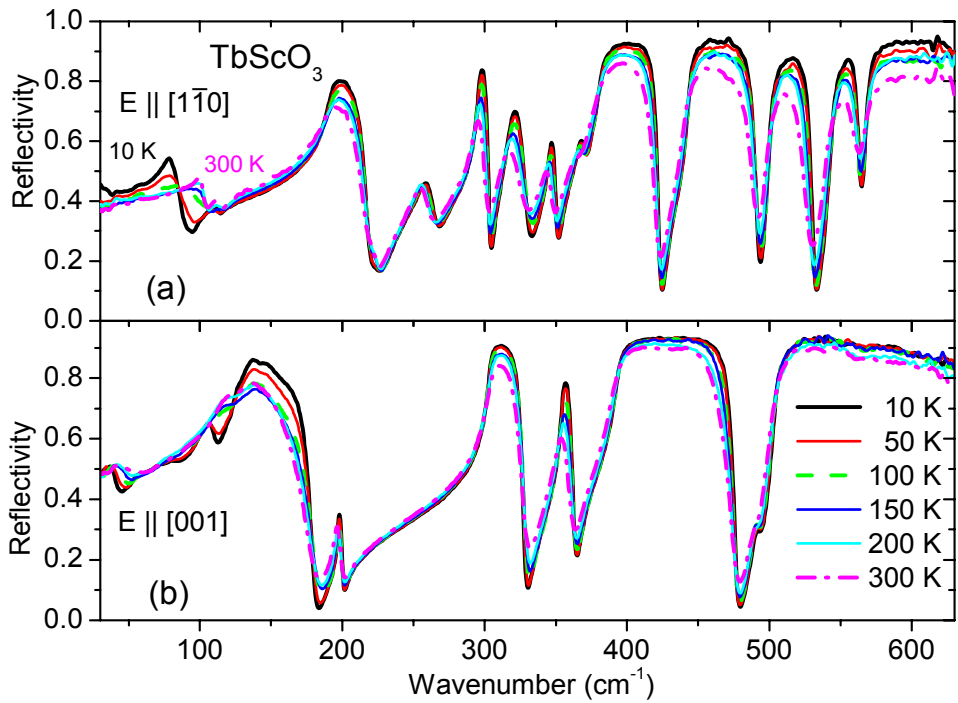

Figure 2. IR reflectivity spectra of a TbScO$_3$ single crystal for polarizations (a) $E \parallel [1\bar{1}0]$ and (b) $E \parallel [001]$ taken at selected temperatures.

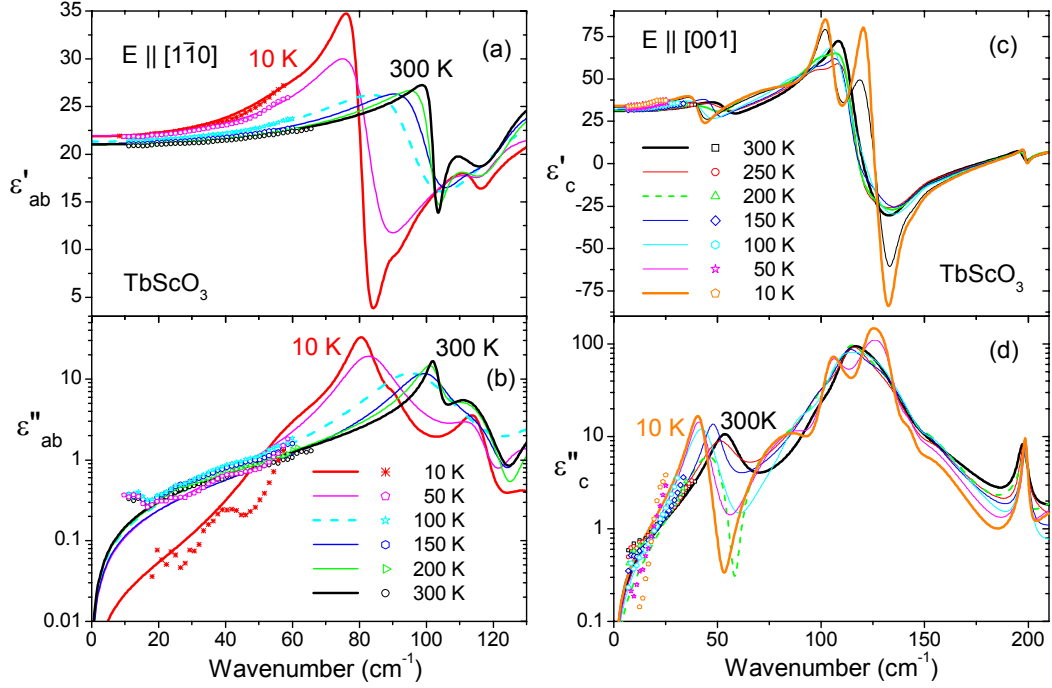

Figure 3. Complex dielectric spectra obtained from fits of IR reflectivity spectra with polarization (a,b) $E \parallel [1\bar{1}0]$ and (c,d) $E \parallel [001]$. Experimental THz dielectric spectra are plotted with symbols. Only the low-frequency region, where changes in the phonon spectra occur with temperature, are plotted.

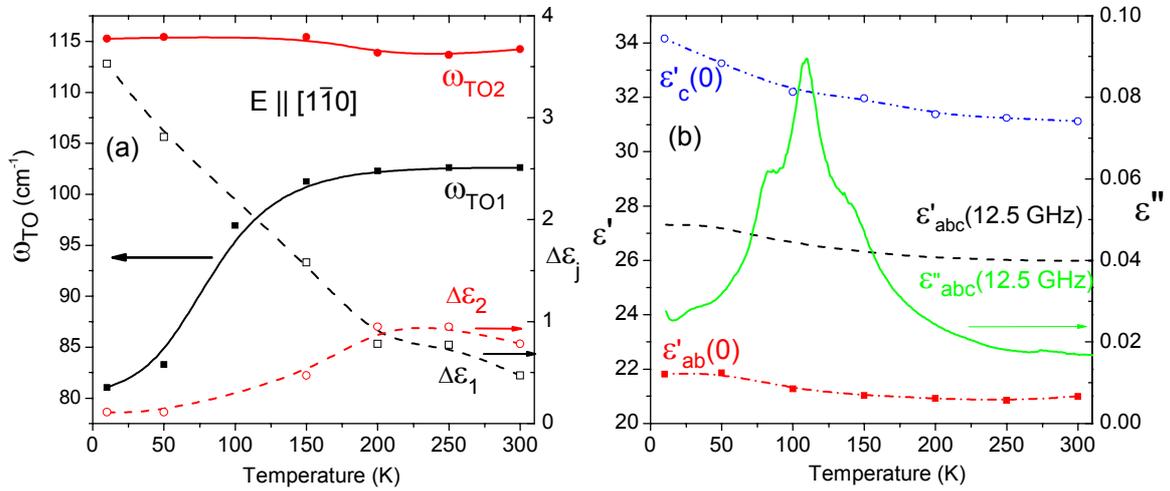

Figure 4. (a) Temperature dependence of the phonon frequencies and dielectric strengths $\Delta\varepsilon_j$ of the two lowest frequency phonons active in $E \parallel [1\bar{1}0]$ polarized spectra. (b) Temperature dependence of static dielectric constant along the c crystal direction ($\varepsilon'_c(0)$) and perpendicular to the c direction ($\varepsilon'_{ab}(0)$) obtained from the sum of phonon and electronic contributions (Eqs. (3) and (4)) compared with measurements of the dielectric constant

measured in the $(1\bar{1}0)$ crystal plane ($\varepsilon'_{abc}$) at 12.5 GHz. Microwave dielectric loss $\varepsilon''_{abc}$ is also shown on the scale on the right axis. The lines connecting points are guides for the eye.

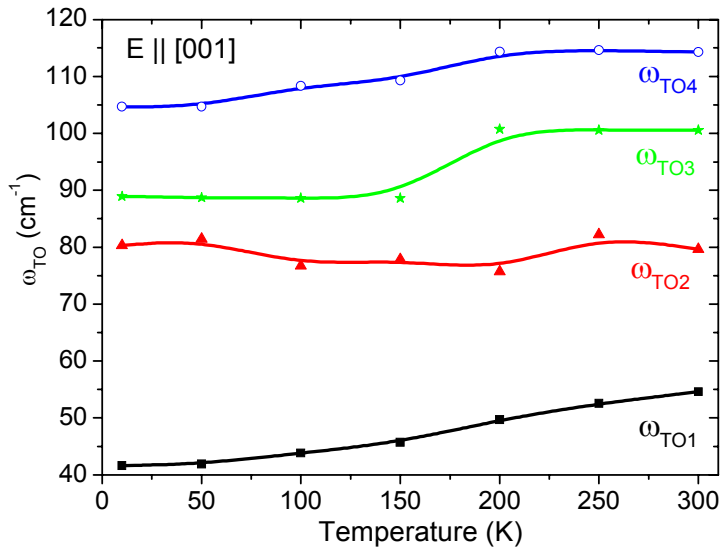

Figure 5. Temperature dependence of the phonon frequencies seen in low-frequency polarized IR spectra with **E** ǁ [001]. The lines connecting the points are guides for the eye.